\documentclass[aps,prb,reprint,twocolumn,superscriptaddress]{revtex4-2}
\usepackage{graphicx}
\usepackage{amsmath}
\usepackage{amsfonts}
\usepackage{color}
\usepackage{amssymb}%
\usepackage{siunitx}
\usepackage{blindtext}

\usepackage{xspace}
\usepackage{siunitx} 
\usepackage{miller}

\usepackage[version=3]{mhchem}
\usepackage{natbib}
\usepackage[]{hyperref}

\providecommand{\U}[1]{\protect\rule{.1in}{.1in}}

\newcommand{\Angrz}{${\AA}^{-1}$}

\newcommand{\sro}{Sr$_2$RuO$_4$}

\newcommand{\uev}[1]{\SI{#1}{\micro\eV}}

\begin{document}
\title{Neutron scattering studies on spin fluctuations in Sr$_{2}$RuO$_{4}$}

\author{K. Jenni}
\email{[e-mail: ]jenni@ph2.uni-koeln.de}
\affiliation{$I\hspace{-.1em}I$. Physikalisches Institut,
Universit\"at zu K\"oln, Z\"ulpicher Str. 77, D-50937 K\"oln,
Germany}

\author{S. Kunkem\"oller}
\affiliation{$I\hspace{-.1em}I$. Physikalisches Institut,
Universit\"at zu K\"oln, Z\"ulpicher Str. 77, D-50937 K\"oln,
Germany}

\author{P. Steffens}
\affiliation{Institut Laue Langevin,71 avenue des Martyrs, 38000 Grenoble, France}

\author{Y. Sidis}
\affiliation{Laboratoire L\'eon Brillouin, C.E.A./C.N.R.S., F-91191 Gif-sur-Yvette CEDEX, France}

\author{R. Bewley}
\affiliation{ISIS, United Kingdom}

\author{Z. Q. Mao}
\affiliation{Department of Physics, Graduate School of Science, Kyoto University, Kyoto
606-8502, Japan}
\affiliation{Department of Physics, Tulane University, New Orleans, LA 70118, USA}
\affiliation{Department of Physics, Pennsylvania State University, University Park, PA 16802, USA}

\author{Y. Maeno}
\affiliation{Department of Physics, Graduate School of Science, Kyoto University, Kyoto
606-8502, Japan}

\author{M. Braden}
\email{[e-mail: ]braden@ph2.uni-koeln.de}
\affiliation{$I\hspace{-.1em}I$. Physikalisches Institut, Universit\"at zu K\"oln,
Z\"ulpicher Str. 77, D-50937 K\"oln, Germany}

\date{\today}

\begin{abstract}
The magnetic excitations in Sr$_2$RuO$_4$ are studied by polarized and unpolarized neutron scattering
experiments as a function of temperature. At the scattering vector of the Fermi-surface nesting with a half-integer out-of-plane component,
there is no evidence for the appearance of a resonance excitation in the superconducting phase.
The body of existing data indicates weakening of the scattered
intensity in the nesting spectrum  to occur at very low energies.
The nesting signal persists up to 290\,K but is strongly reduced.
In contrast, a quasi-ferromagnetic contribution maintains its strength and still exhibits a finite width
in momentum space.

\end{abstract}

\pacs{7*******}
\maketitle











\section{introduction}

A quarter century after the discovery of superconductivity in \sro \cite{Maeno1994} its character and
its pairing mechanism remain mysterious. Inspired by the ferromagnetic order appearing in the metallic
sister compound SrRuO$_3$ \cite{Koster2012} it was initially proposed that ferromagnetic fluctuations drive the superconductivity
in \sro \ rendering its superconductivity similar to the A-phase of superfluid ${}^{3}$He  \cite{Baskaran1996,Rice1995}.
For a long time chiral $p$-wave superconductivity with spin-triplet pairing has been considered to best describe
the majority of experimental studies \cite{Mackenzie2003,Maeno2012} although the absence of detectable
edge currents \cite{Hicks2010} and the constant Knight-shift observed for fields perpendicular to the Ru layers \cite{Murakawa2004}
were not easily explained in this scenario \cite{Kallin2012}. Further insight was gained from experiments performed under large
uniaxial strain that revealed a considerable enhancement of the superconducting transition temperature by more than a factor two \cite{Hicks2014,Steppke2017}, similar to the enhancement in the eutectic crystals \cite{Maeno2012}.
However, the breaking of the four-fold axis should split the superconducting transition of the chiral state in contradiction
with a single anomaly appearing in the specific heat under strain \cite{Li2019}. Furthermore, the strain dependence of the transition
temperature close to zero strain is flat \cite{Hicks2014,Barber2019}, whereas one expects a linear dependence for the chiral state.

The picture of chiral $p$-wave superconductivity was fully shaken, when the two experiments yielding the strongest support for triplet pairing
\cite{Ishida1998,Duffy2000} were revised. The new studies of the Knight shift in NMR \cite{Pustogow2019,Ishida2020} and those of the
polarized neutron diffraction \cite{Petsch2020} reveal an unambiguous drop of the electronic susceptibility that is inconsistent
with spin-triplet pairs parallel to Ru layers. Since then, numerous proposals for the superconducting state were made mostly
invoking some $d$-wave state and the discussion of the superconducting pairing has become very active \cite{Roising2019,Romer2019,Wang2019,Wang2020,Suh2020,Romer2020,Kivelson2020,Sharma2020}.
The observations of broken time-reversal symmetry in muon spin relaxation experiments \cite{Luke1998,Grinenko2020} and in measurements
of the magneto-optical Kerr effect \cite{Xia2006} may require interpretations other than the chiral $p$-wave scenario. Many theories discuss a
superconducting state with a complex combination of components \cite{Roising2019,Romer2019,Wang2019,Wang2020,Suh2020,Romer2020,Kivelson2020}.

Assuming a simple boson-mediated pairing following BCS theory, phonons and magnetic fluctuations
or a combination of both \cite{Schnell2006} can be relevant. There are anomalies in the phonon dispersion that could be fingerprints
of electron phonon coupling \cite{Braden1997,Braden2007}. The phonon mode that describes the rotation of the RuO$_6$ octahedra around the
$c$ axis exhibits an anomalous temperature dependence and severe broadening \cite{Braden1997}. This mode can be associated with the
structural phase transition and with the shift of the van Hove singularity in the $\gamma$ band through the Fermi level.
Both effects occur upon small Ca substitution \cite{Fang2001,Friedt2001}.
In addition, the Ru-O bond-stretching modes that exhibit an anomalous downwards dispersion in many oxides with
perovskite-related structure \cite{Braden2002a} exhibit an anomalous dispersion in \sro \ as well \cite{Braden2007}.
Comparing the first-principles calculated \cite{Wang2010} and measured \cite{Braden2007} phonon dispersion
in \sro \ the agreement is worst for these longitudinal bond-stretching modes, which exhibit a flatter dispersion indicating better
screening compared to the density functional theory (DFT) calculations.
Note, however, that perovskite oxides close to charge ordering exhibit a much stronger renormalization of the zone-boundary modes with
breathing character that is frequently labeled overscreening \cite{Braden2002a,Braden2005}.

On the other side there is clear evidence for strong magnetic fluctuations deduced from NMR \cite{Imai1998} and inelastic neutron scattering (INS) experiments \cite{Sidis1999,Braden2002,Servant2002,Braden2004,Iida2011,Iida2012,Steffens2019}.
The dominating magnetic signal is incommensurate and stems from nesting in the one-dimensional bands associated with $d_{xz}$ and $d_{yz}$ orbitals, see Fig. 1.
The relevance of this instability towards an incommensurate spin-density wave (SDW) is underlined by the observation of static magnetic order emerging at this $\bf q$ position in reciprocal space for minor substitution of Ru by Ti \cite{Braden2002b} or of Sr by Ca \cite{Carlo2012,Kunkemoeller2014}.
A repulsive impurity potential was recently proposed to form the nucleation center for the magnetic ordering that should strongly
couple to charge currents \cite{Zinkl2020}.
Furthermore, the temperature dependence of these incommensurate magnetic fluctuations
in pure \sro \ agrees with a closeness to a quantum critical point \cite{Braden2002}. These nesting-induced magnetic fluctuations can easily be explained by DFT
calculations using the random phase approximations (RPA) \cite{Mazin1999} but their relevance for the superconducting pairing remains controversial \cite{Huo2013}.
Inelastic neutron scattering in the superconducting state can exclude the opening of a large gap for these nesting-driven \cite{Kunkemoeller2017}. Since magnetic
excitations are particle hole excitations one expects in the most simple isotropic case a magnetic gap comparable to twice the superconducting one, which can be safely excluded.
However, the anisotropy of the gap function and interactions can strongly modify the magnetic response in the superconducting state.
A more recent TOF inelastic neutron scattering experiments confirms the absence of a large gap but reports weak evidence for suppression of spectral weight at
very low energies \cite{Iida2020}. This experiment also claims the occurrence of a spin resonance mode at the nesting position with a finite perpendicular wave-vector component,
which would point to an essential modulation of the superconducting gap perpendicular to the RuO$_2$ layers but which is inconsistent with the results of this work.

In addition to the incommensurate nesting-induced fluctuations,
macroscopic susceptibility \cite{Maeno1997}, NMR \cite{Imai1998,Mukuda1999} and  also polarized inelastic neutron scattering experiments \cite{Braden2004,Steffens2019}
reveal the existence of magnetic fluctuations centered at the origin of the Brillouin zone, which typically can be associated with ferromagnetism.
Futhermore, a small concentration Co doping can lead to static short-range ferromagnetic order \cite{Ortmann2013}.
All techniques find almost temperature independent quasi-ferromagnetic excitations in pure Sr$_2$RuO$_4$. This ferromagnetic response
qualitatively agrees with a recent dynamical mean field theory (DMFT) analysis of magnetic fluctuations \cite{Strand2019},
which finds essentially local magnetic fluctuations superposed on the well known nesting signal.
However, the neutron data disagree with a fully local character as they show a finite $q$ dependence \cite{Steffens2019}.
The quasi-ferromagnetic fluctuations also disagree with the expectations for a nearly ferromagnetic system that
exhibits paramagnon scattering \cite{Moriya1985,Steffens2019}.
SrRuO$_3$ clearly exhibits such paramagnon scattering with its
well-defined structure in $q$ and energy space \cite{Jenni2019}.

Here we present additional neutron scattering experiments on the magnetic fluctuations in \sro , which
focus on several aspects that are particularly relevant for the superconducting pairing mechanism involving magnetic fluctuations or for the general
understanding of magnetic excitations in a strongly correlated electron system.
We discuss the possibility of important out-of-plane dispersion in the magnetic response in the superconducting and normal
states, the shape of nesting scattering away from the peak position and the non-local character of the quasi-ferromagnetic response.


\section{Experimental}

\begin{figure}
 \includegraphics[width=0.75\columnwidth]{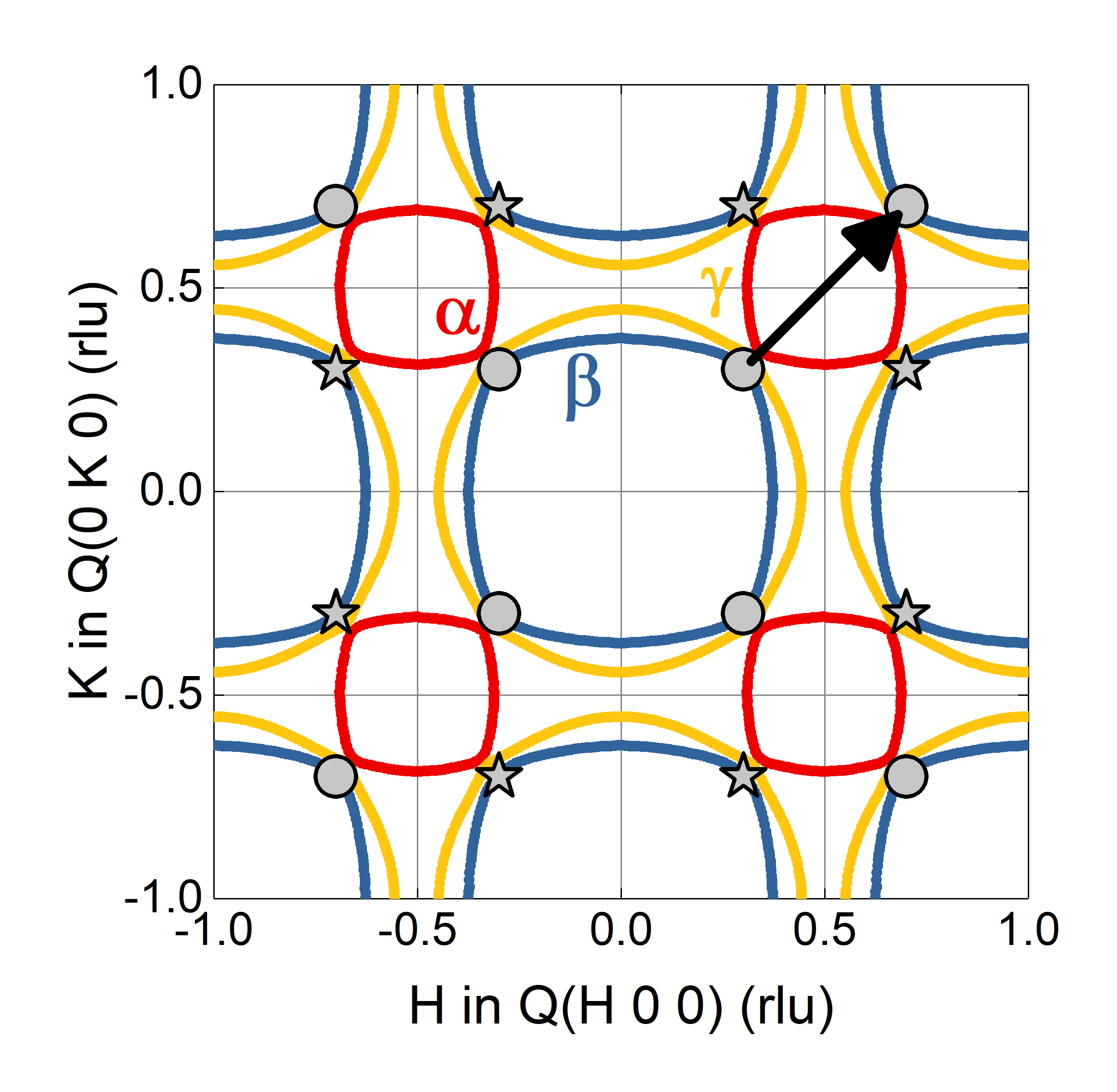}
  \caption{Fermi surface of \sro\ for $k_z=0$. The bands are based on LDA+SO calculations from \cite{Veenstra2013} and marked by different colors. The black arrow represents the dominant nesting vector between the one-dimensional sheets $\alpha$ (red) and $\beta$ (blue). The incommensurate positions of the in-plane nesting signal are marked by different symbols. The circles represent the crystallographically equivalent positions ($\pm$0.3,$\pm$0.3) and ($\pm$0.7,$\pm$0.7). The positions ($\pm$0.3,$\pm$0.7) and ($\pm$0.7,$\pm$0.3) shown by stars are equivalent to those on the diagonals only in a purely two-dimensional picture, because (1 0 0) is not an allowed Bragg peak in the body-centered lattice.
  }
  \label{Fermisurface}
 \end{figure}

INS experiments were carried out on the ThALES \cite{THALES2018,THALES2020} and IN20 \cite{IN2020} triple-axis spectrometers (TAS) at the Institut Laue Langevin and on the LET \cite{LET2017} time-of-flight (TOF) spectrometer at the ISIS Neutron and Muon Source. We used an assembly of 12 \sro \ crystals with a total volume of 2.2 cm$^3$ in all experiments.
At Kyoto University, the crystals were grown using the floating zone method and similar crystals were studied in many experiments \cite{Mackenzie2003,Maeno2012}. The crystal assembly was oriented in the [100]/[010] scattering plane (corresponding to a vertical $c$ axis) to study the in-plane physics of the Ru layers. Additionally, with the instruments ThALES and LET it was possible to access parts of the $q$ space perpendicular to the plane which enables an analysis of the out-of-plane dispersion of the magnetic response.
To conduct experiments inside the superconducting phase a dilution refrigerator was used, reaching a temperature of $\sim$200\,mK, well below the transition temperature of $\sim$1.5\,K.
ThALES and LET are operating with a cold neutron source providing the energy resolution to study the magnetic response down to $\sim$\uev{200}. The TOF spectrometer LET records data simultaneously with four different values of the incidental energies, $E_i$ and resolutions, while the energy resolution of the TAS ThALES is determined by the chosen final neutron wave vector $k_f$ of 1.57\,\Angrz \ combined with the collimations. On ThALES the best intensity to background ratio was achieved by using a  Si(111) monochromator and PG(002) analyzer combined with a radial collimator in front of the analyzer for further background reduction.
The same configuration was also used in an earlier study \cite{Kunkemoeller2017}.

A polarized neutron scattering experiment was performed on the thermal TAS IN20 using Heusler crystals as monochromator and analyzer. A spin flipper in front of the analyzer enabled the polarization analysis. The scans were performed with a fixed final momentum of $k_f=$ 4.1\,\Angrz , where the graphite filter in front of the analyzer cuts higher order contaminations. Longitudinal polarization analysis was performed with a set of Helmholtz coils.

\section{Results and discussion}

\begin{figure}
 \includegraphics[width=\columnwidth]{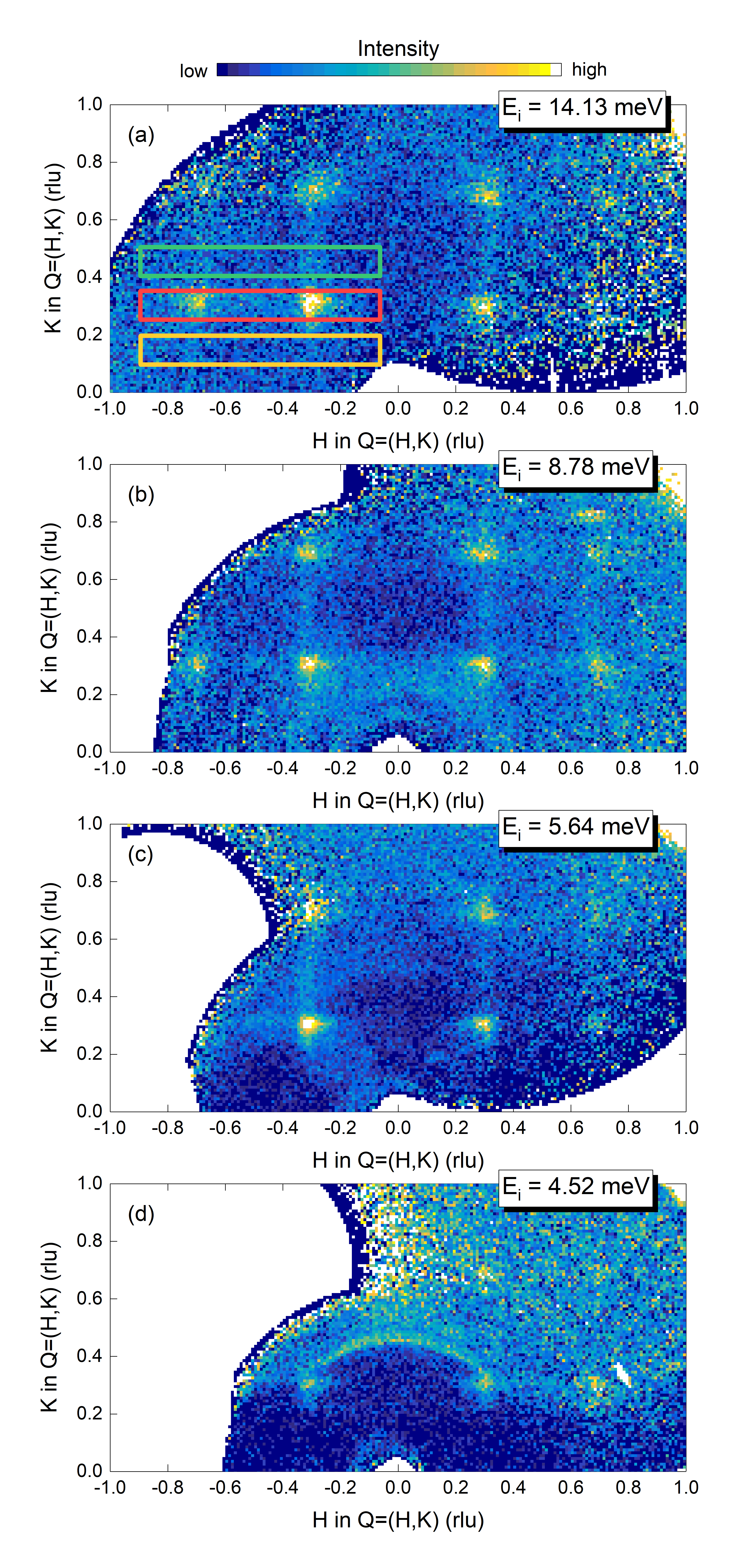}
  \caption{In-plane scattering in the superconducting phase (T = 0.2\,K). The TOF data at four different incidental energies display the magnetic scattering distribution in the $ab$ plane. The intense signal at the incommensurate positions (0.3,0.3), (0.7,0.7), and (0.3,0.7) is visible for all E$_i$. Additionally there is magnetic scattering between the incommensurate positions in [$\xi$,0] and [0,$\xi$] directions, respectively. To increase the statistics the data is integrated over the maximum $L$ range of [-0.7,0.7] and full $E$ range depending on the incidental energy (1.75$<E<$10 for $E_i=14.13 \mathrm{meV}$, 0.8$<E<$6.7 for $E_i=8.78 \mathrm{meV}$, 0.7$<E<$4.5 for $E_i=5.64 \mathrm{meV}$, and 0.5$<E<$3.5 for $E_i=4.52 \mathrm{meV}$). The overlayed rectangles  in (a) represent the integration area of the one-dimensional cuts displayed in Fig. \ref{MapCuts}(a)-(c).
  }
  \label{Qmap}
 \end{figure}

\subsection{$q$ dependence of fluctuations associated with nesting}
The TOF technique enables an imaging of the complete $\mathbf{Q}$-$E$-space, which gives insight on the distribution of scattering intensity in the reciprocal space.
Throughout the paper, the scattering vector $\mathbf{Q}$=($H$,$K$,$L$) and the propagation vector in the
first Brillouin zone $\mathbf{q}$=($q_h$,$q_k$,$q_l$) are given in reciprocal lattice units (rlu).
We mostly consider only the planar wave vector $\mathbf{Q_{2d}}$=($H$,$K$) projection.
Fig.  \ref{Qmap} shows the inelastic scattering plotted against the $H$,$K$ components of the scattering vector in the superconducting phase. The four different panels display sections of the two-dimensional ($H$,$K$) plane for different incident energies and hence different resolutions. The intensities are fully integrated along the energy transfer (depending on the incident energy) and along  the out-of-plane component of the scattering vector, -0.7$<L<$0.7. The high scattering intensities at the incommensurate positions ($\pm$0.3,0.3), ($\pm$0.3,0.7), and ($\pm$0.7,0.7) are clearly visible, arising from the well known antiferromagnetic fluctuations \cite{Sidis1999,Servant2002,Braden2002,Iida2011,Iida2012}. Additionally there are ridges of scattering intensities connecting these positions in [$\xi$,0] and [0,$\xi$] directions that were first reported in \cite{Iida2011,Iida2012}.
The arc visible in Fig.\,\ref{Qmap}(d) connecting (-0.3,0.3) and (0.3,0.3) is a spurious signal; it does not appear for the other incidental energies.

Neglecting electronic dispersion perpendicular to the planes and assuming an idealized scheme of flat one-dimensional bands originating from the $d_{xz}$ and $d_{yz}$ orbitals, one expects nesting induced magnetic excitations for any two-dimensional vector $\bf Q_{2d}$=(0.3,$\xi$) and ($\xi$,0.3) and accordingly a peak at (0.3,0.3) \cite{Mazin1999}. The peaks clearly dominate but the ridges are also detectable - mostly for the positions connecting the nesting peaks, i.e. 0.3$<\xi<$0.7. This is in accordance with the calculation of the bare susceptibility which shows an enhanced signal only between the peaks, i.e. for the paths from (0.3,0.3) to (0.7,0.3) \cite{Mazin1999}.

\begin{figure}
 \includegraphics[width=\columnwidth]{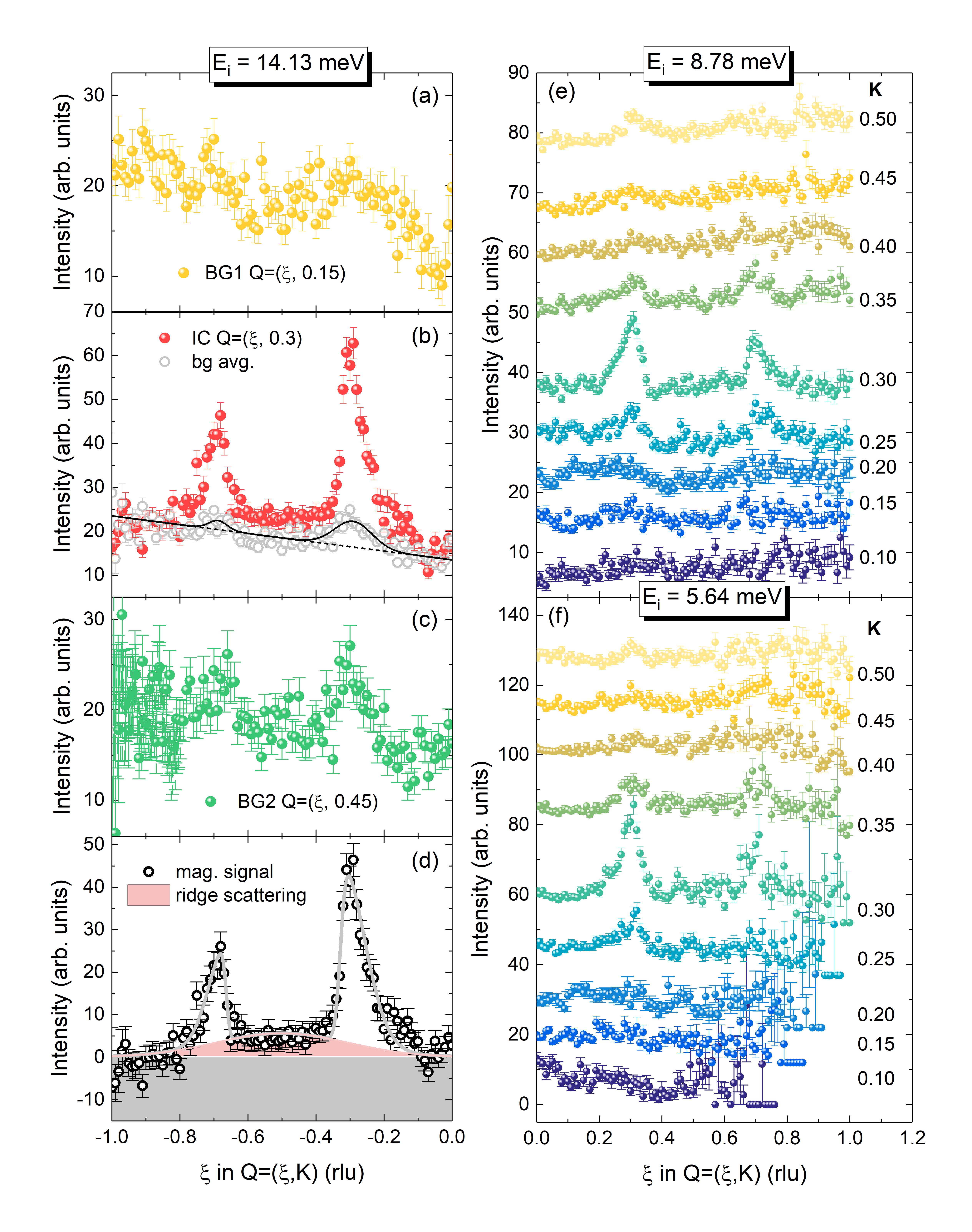}
  \caption{Magnetic scattering along the connection of the incommensurate positions. (a)-(c) show one-dimensional cuts from Fig.  \ref{Qmap}(a) along the ($\xi$,$K$) paths for $K$ = 0.15 (a), 0.3 (b), and 0.45 (c). The background at both sides of the incommensurate positions is displayed in (a) and (c) (represented by the same colored rectangles in \ref{MapCuts}(a)). An averaged background is formed from both (gray open circles) and fitted with a linear contribution and two Gaussians (black solid line). This is compared to the incommensurate signal in (b). In (d) the linear background contribution (black dashed line) is subtracted and the signal along the [$\xi$,0] direction is fitted with two skew Gaussians for the incommensurate signal and a broad Gaussian fixed at $\xi=0.5$ (red area) taking into account the ridge scattering. (e) and (f) represent one-dimensional cuts for different $K$ and two different incident energies 8.78\,meV and 5.64\,meV taken from Fig. \ref{Qmap}(b) and (c). The integration range in [0,$\xi$] direction is $\pm0.025$ around the $K$ value and the scans are shifted vertically for better visibility.
  }
  \label{MapCuts}
 \end{figure}

To analyze the ridge scattering and the anisotropy of the incommensurate signals in detail, Fig. \ref{MapCuts}(a) shows one-dimensional cuts along the ridge in [$\xi$,0] direction calculated from the data taken with $E_i$=14.13\,meV (Fig.  \ref{Qmap}(a)). By subtracting the background obtained from the average of ($\xi$,0.15) and ($\xi$,0.45), shown in Fig. \ref{MapCuts}(b) and (c) respectively, we isolate the signal along the line ($\xi$,0.3) shown in Fig. \ref{MapCuts}(d). The ridge scattering is mainly detectable between the peaks at the incommensurate positions, as it is visible in the two one-dimensional cuts representing the background parallel to the ridge on both sides (Fig. \ref{MapCuts}(b) and (c)).
While the ($\xi$,0.15) cut exhibits only a weak signal around (-0.3,0.15) the ($\xi$,0.45) cut shows clearly two peaks at the (-0.7,0.45) and (-0.3,0.45) positions representing the ridges in [0,$\xi$] direction. The rounding of the one-dimensional Fermi-surface sheets suppresses the susceptibility at (0.3,$\xi$) with $\xi$ lower than 0.3, but this suppression is not abrupt.
Besides the ridge scattering we may also confirm the pronounced asymmetry of the nesting peak with a shoulder near (0.25,0.3) and equivalent positions. This shoulder was reported in \cite{Braden2002} and was also found in the full RPA calculations.

The asymmetry of the nesting peaks and the ridge scattering between the incommensurate positions can also be seen in the data of lower incident energies (see Fig. \ref{MapCuts}(e) and (f)). The one-dimensional cuts for different $K$ values confirm the asymmetric shape of the nesting peaks.
A thorough analysis of the pure magnetic signal as in the case of $E_i=14.13$\,meV is not possible due to uncertainty in the background. Furthermore, the ridge scattering is less pronounced in the data obtained with lower incident energies, which indicates a higher characteristic energy of the ridge scattering.
This further explains why the much weaker scattering in the ridges has not been detected in early TAS studies \cite{Sidis1999,Servant2002,Braden2002}.

\subsection{Search for gap opening or a resonance mode below $T_c$}
\begin{figure}
 \includegraphics[width=\columnwidth]{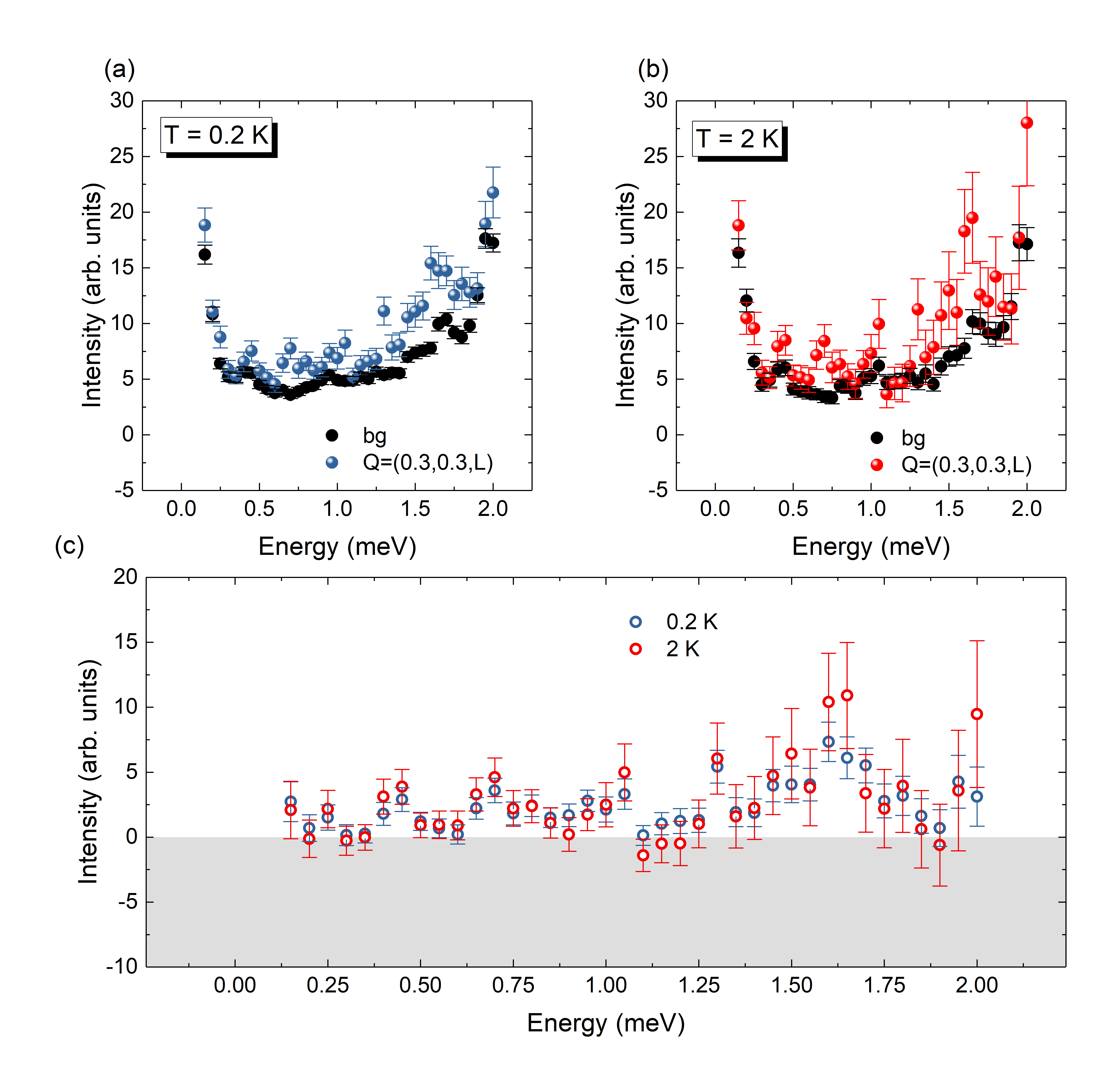}
  \caption{Low energy dependence of incommensurate signal below and above the superconducting transition extracted from TOF data. (a) and (b) display the energy scans at  at $q$=(0.3,0.3) below (T = 0.2\,K) and above (T = 2\,K) the superconducting phase transition. The background in both panels is derived from the constant $Q$ cut at (0.09,0.41) for both temperatures ($|\textbf{Q}_{IC}|=|\textbf{Q}_{bg}|$). To increase statistics the TOF data with an incidental energy of 3\,meV is fully integrated over $L$ (range [-0.7,0.7]) and symmetrized by folding in $q$ space at (0.3,0.7) along the (1,-1,0) plane. The H and K component is integrated with the range [0.25,0.35] (c) Background subtraction and Bose factor correction yields the pure magnetic response at low energies which is compared inside and outside the superconducting phase.
  }
  \label{EscansTOF}
 \end{figure}

The opening of a superconductivity-induced gap in the spectrum of magnetic fluctuations would have strong impact on the discussion of the superconducting character in \sro .
Previous INS experiments using a TAS revealed the clear absence of a large gap at the nesting position \cite{Kunkemoeller2017}, whereas a recent TOF experiment reports a tiny gap although the statistics
remained very poor \cite{Iida2020}. Studying the magnetic response of \sro \ in its superconducting phase by INS is challenging, because one needs to focus on small energies of the order of 0.2 to 0.5\,meV.
At these energies the signal in the normal state is at least one order of magnitude below its maximum strength at 6 meV, and the required high-energy resolution further suppresses
statistics.
Fig.  \ref{EscansTOF} presents the TOF data obtained with $E_i=3$\,meV by calculating the energy dependence at the nesting position integrated over all $L$ values.
The full $L$ integration is needed to enhance the statistics.
In Fig.  \ref{EscansTOF} (a) and (b) we compare the raw data for both temperatures with the background signal. In (c) the background subtracted magnetic response in the superconducting phase is compared to that in the normal phase. There is no evidence for the opening of a gap within the statistics of this TOF experiment.
Also a resonance at a finite energy cannot be detected. Admittedly the statistics of this TOF data is too poor to detect small signals or their suppression.

\begin{figure}
 \includegraphics[width=0.8\columnwidth]{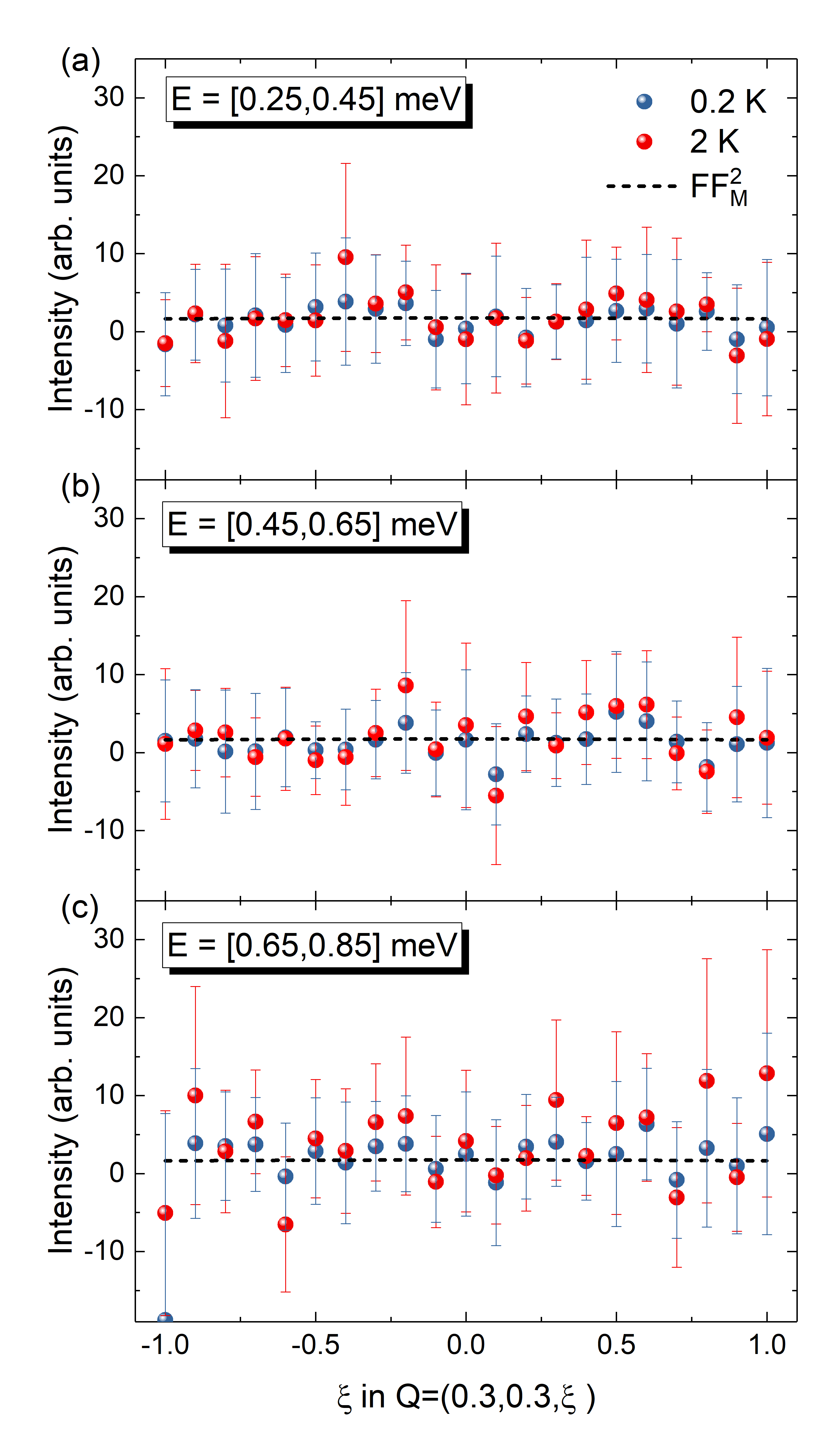}
  \caption{L dependence of incommensurate signal at low energies extracted from TOF data. Constant E cuts with an integration width of 0.2\,meV at the incommensurate position (0.3,0.3,$L$) were adjusted for the measured background at the same energy at (0.09,0.41,$L$) and corrected for the Bose factor. The $L$ dependence of the magnetic response in the superconducting phase (blue) is compared to the normal phase (red). Additionally, the square of the Ru$^{1+}$ form factor is depicted in each panel (black dashed line). There is no evidence for a peak at $L$=0.5.
  }
  \label{LscansTOF}
 \end{figure}

Following the claim of Iida et al. \cite{Iida2020} the TOF data is also analyzed in terms of a possible resonance mode appearing at a finite value of the $L$ component, i.e. at (0.3,0.3,0.5). Therefore, the $L$ dependence of the magnetic signal at (0.3,0.3,$L$) is determined by background subtraction and compared for the two temperatures (see Fig. \ref{LscansTOF}). The different panels represent the  energy ranges from reference \onlinecite{Iida2020}, where a resonance appearing at 0.56 meV is proposed for $L$=0.5. In our data shown in Fig.  \ref{LscansTOF}(b), there is no difference visible between superconducting and normal phase at $L$=$\pm$0.5.

\begin{figure}[htbp]
 \includegraphics[width=0.8\columnwidth]{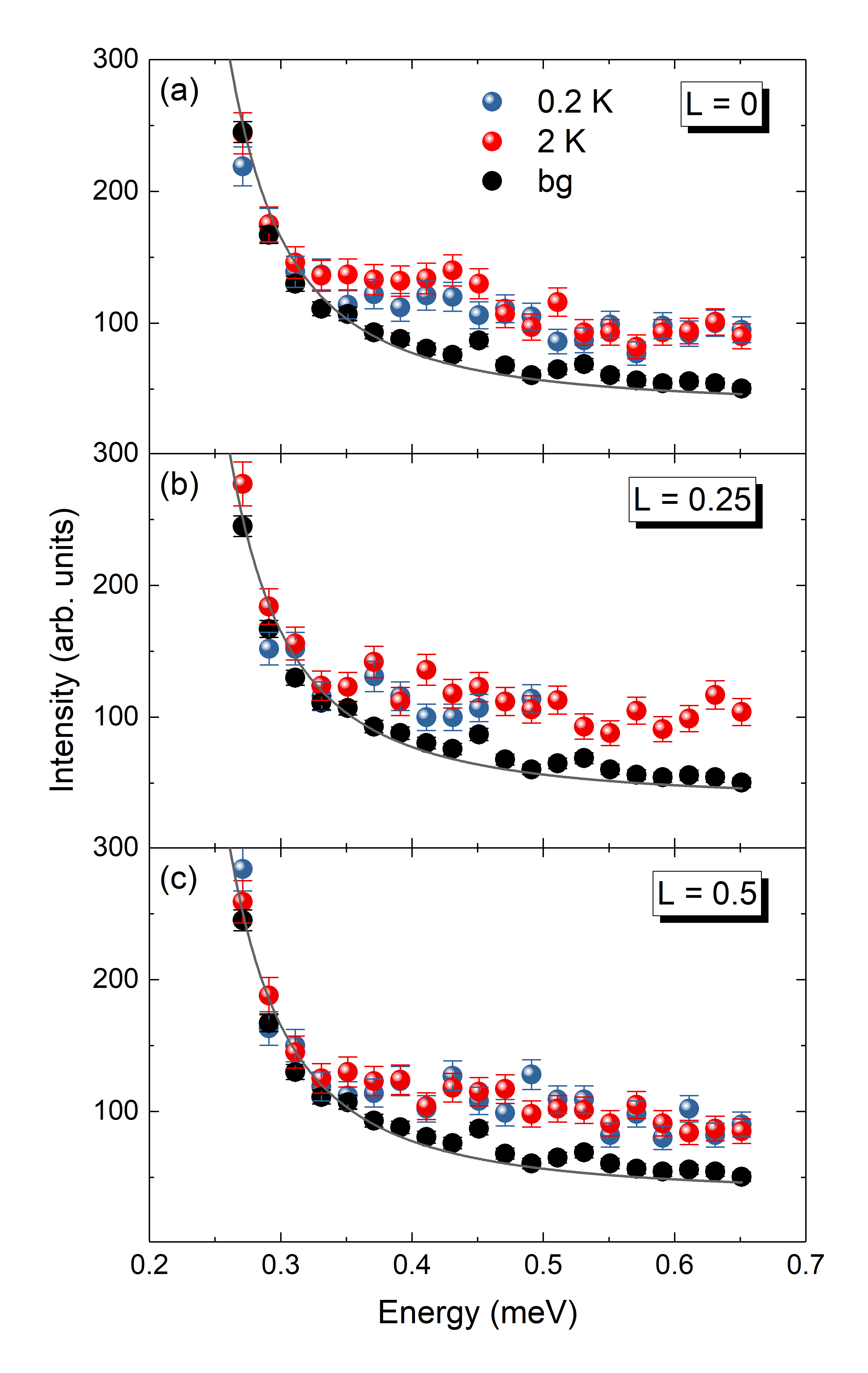}
  \caption{L dependence of incommensurate signal at low energies extracted from TAS data. The constant q scans were conducted at the incommensurate positions (0.3,0.7,$L$) with $L$ = 0, 0.25, and 0.5 in the superconducting and normal phase. The background for each $L$ is measured after $\omega$ rotation of $20\deg$, thus keeping $|q|$ constant, and later averaged for all scans, yielding the presented background (black circles) and its fit (gray). The intensity is normalized with 1980000 monitor counts which
corresponds to a measuring time of about 15 min per point.
  }
  \label{LdepTAS}
 \end{figure}

\begin{figure}
 \includegraphics[width=\columnwidth]{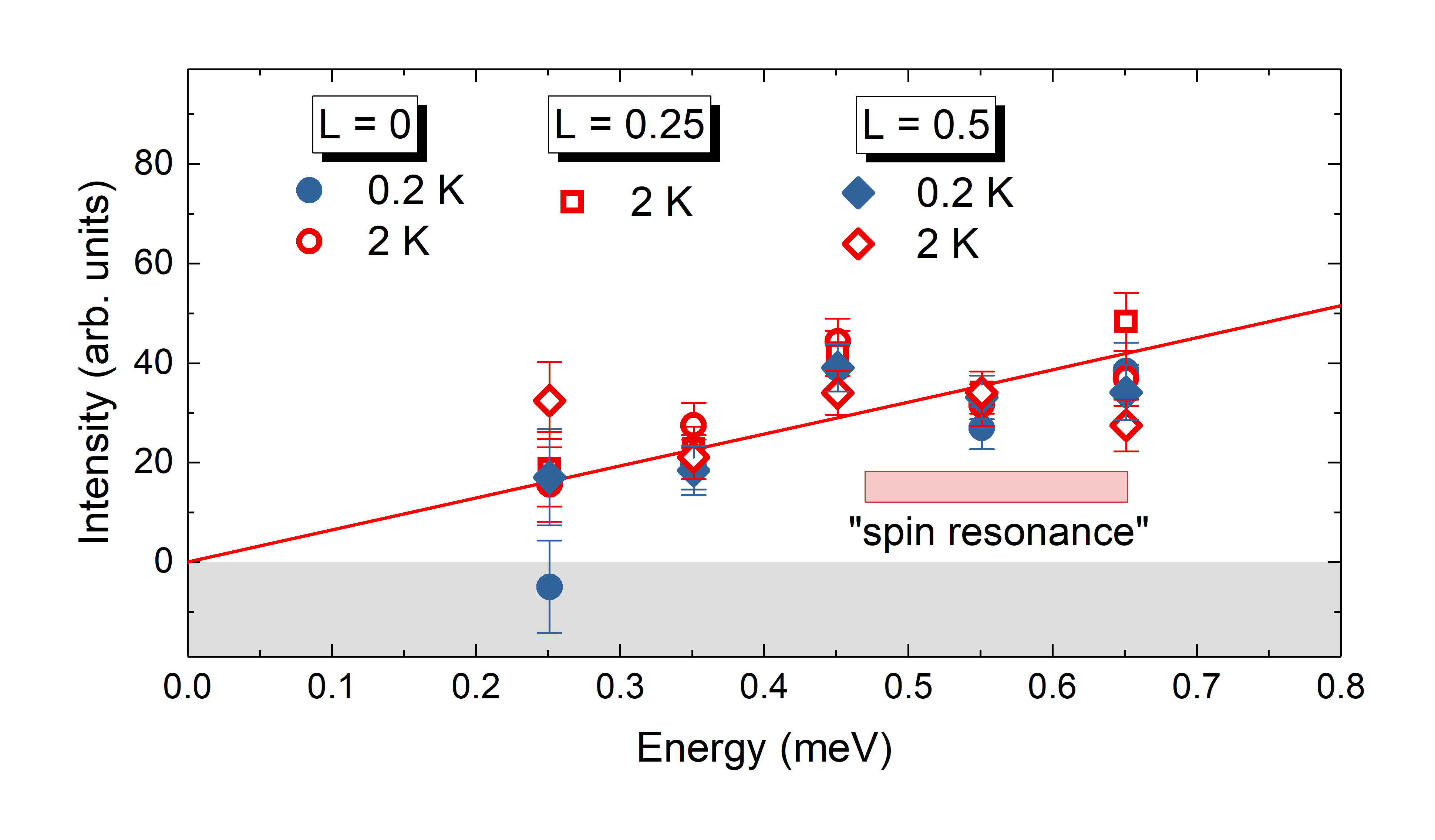}
  \caption{Comparison of the background-free incommensurate signal Q=(0.3,0.7,$L$) for different $L$ values and temperatures. The compared data originates from the constant $q$ scans shown in Fig.  \ref{LdepTAS}. The binning is increased to $\Delta$ E = 0.1\,meV which yields better statistics. A linear fit (red line) provides a guide to the eye. The energy range of the proposed spin resonance \cite{Iida2020} is indicated by the red box.
  }
  \label{compareLTAS}
 \end{figure}

To study the low-energy response and its $L$ dependence in more detail and with better statistics the TAS is better suited since measurements can be focused to single $\bf Q$,E points. Using ThALES and its high flux and energy resolution constant $\bf Q$ scans at the incommensurate position (0.3,0.7,$L$) with $L$=0, 0.25, and 0.5 were measured to investigate the $L$ dependence of the low energy response (see Fig.  \ref{LdepTAS}). This incommensurate position was chosen due to a better signal-to-noise ratio compared to (0.3,0.3,$L$) and because the larger $|Q|$ value allows one to reach finite $L$ values by tilting the cryostat. Similar to Fig.  \ref{EscansTOF} (a) and (b) the raw data for two temperatures is shown in Fig.  \ref{LdepTAS} (a)-(c). The background was measured by rotating $\omega$ by 20 degrees for each $L$ value and then combining all three backgrounds to an average.
For all $L$ values the intensity of the incommensurate signal increases approximately linearly for small energies, following the established single relaxor behavior.
Comparing the two temperatures there is no difference noticable for any $L$ value down to the energy resolution. Especially around 0.56 meV where Iida et al. \cite{Iida2020} propose a resonance at the incommensurate position (0.3,0.3,0.5) the two temperatures yield comparable signals. It should be noted here that while the incommensurate positions (0.3,0.3,0) and (0.3,0.7,0) are crystallographically not equivalent both positions become equivalent with the $L$ component 0.5, see Fig. 1. Therefore, the data taken at (0.3,0.3,0.5) and (0.7,0.3,0.5) can be compared.
To emphasize the absence of a resonance mode around 0.56 meV the data from Fig. \ref{LdepTAS} is plotted with a larger energy binning to further increase the statistics (see Fig. \ref{compareLTAS}, which also indicates the broad energy integration used in \cite{Iida2020}). There is no significant deviation from the general linear behavior for any $L$ value at low temperatures detectable. Iida et al. \cite{Iida2020} report an increase of signal of $\sim$60\% for $L$=0.5 in the superconducting phase, which clearly is incompatible with our data that offer higher statistics.

\begin{figure}
 \includegraphics[width=\columnwidth]{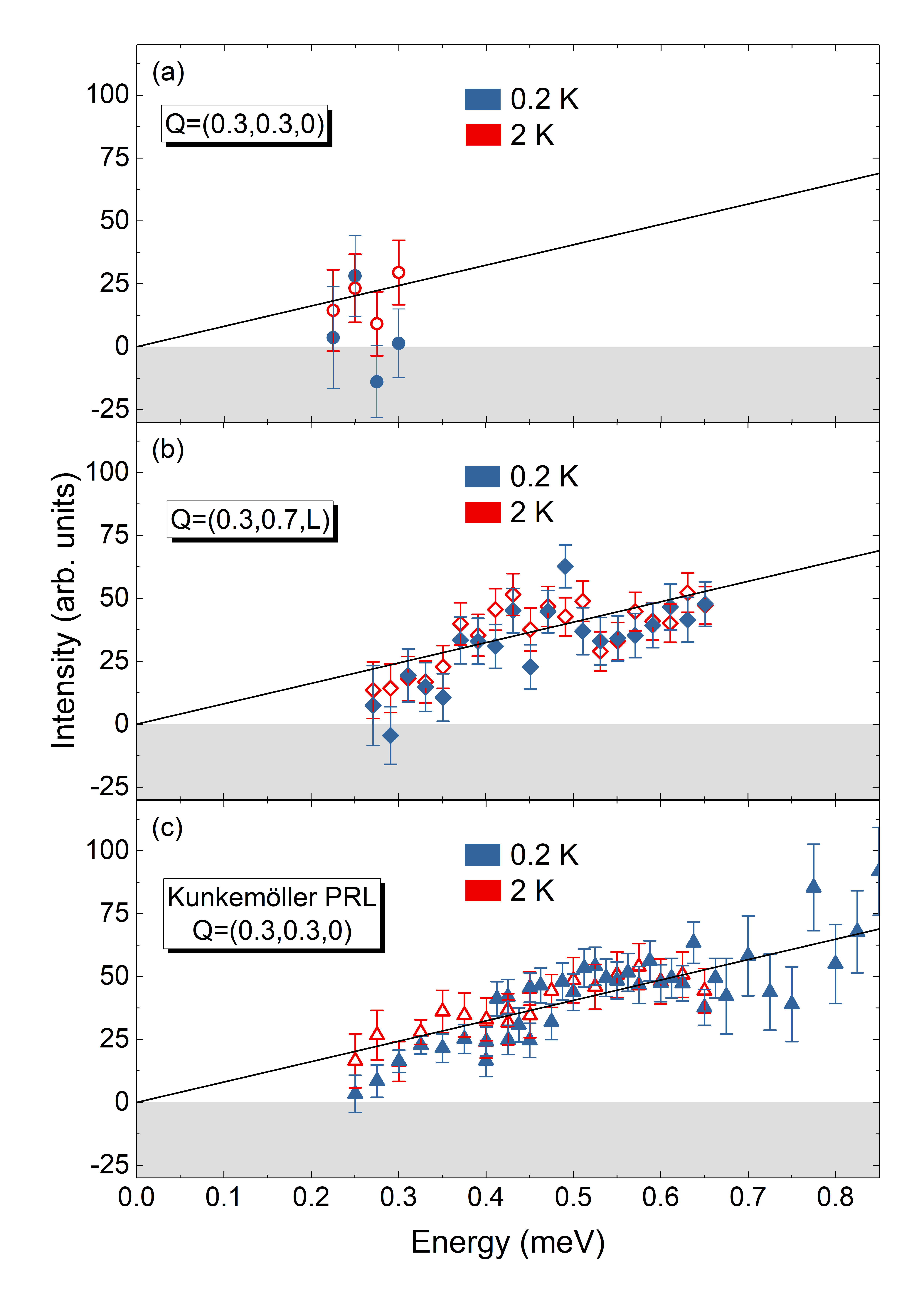}
  \caption{Comparison of the energy dependence of the incommensurate signal with former published data from \cite{Kunkemoeller2017} (labelled: Kunkem\"oller PRL).
  Background corrected data recorded at (0.3, 0.3, 0) (circles) is given in panel (a); the background free signal at (0.3, 0.7) is averaged over all $L$ values for both temperatures (diamonds), panel (b), and the incommensurate signal reported in \cite{Kunkemoeller2017} (triangles) is shown in (c). Data were corrected for the Bose and magnetic form factors.
  }
  \label{gapPRL}
 \end{figure}

Since no $L$ dependence of the magnetic low energy response can be established (Fig.  \ref{LdepTAS} and \ref{compareLTAS}) we merge the data and compare it with the former published low-energy dependence of the incommensurate signal \cite{Kunkemoeller2017} (see Fig.  \ref{gapPRL}). The new experiments below $T_c$ fully confirm that the nesting
excitations in Sr$_2$RuO$_4$ do not exhibit a large gap, i.e. a magnetic gap comparable to twice the superconducting one.
Combining all the previous and new data there is, however, some weak evidence for the suppression of magnetic scattering at very low energies below 0.25\,meV. With the neutron instrumentation of today it seems very difficult to further characterize the suppression of the small signal at such low energy.

For the previously assumed superconducting state detailed theoretical analyzes of the magnetic response were reported \cite{Huo2013}, but
concerning the more recently proposed superconducting symmetries \cite{Roising2019,Romer2019,Wang2019,Wang2020,Suh2020,Romer2020,Kivelson2020,Sharma2020} such investigations lack.
The  $d_{x^2-y^2}$ state deduced from quasiparticle interference imaging \cite{Sharma2020} exhibits nodes at Fermi-surface positions that are connected through the nesting vector.
This implies that even at very low energies, the nesting induced excitations are not fully suppressed in such $d_{x^2-y^2}$  superconducting state, in agreement with the
experimental absence of a large gap in the nesting spectrum \cite{Kunkemoeller2017}.
Within the $d_{x^2-y^2}$ superconducting state the nesting vector also connects Fermi-surface regions with maximum and minimum gap values and
it connects either two regions of the $\beta$ sheet or one  $\beta$ region with an $\alpha$ region. Therefore the conditions for a spin-resonance mode are more
complex and less favorable than in the case of the FeAs-based superconductors, where the $s^{+-}$ superconducting symmetry and the nesting magnetic fluctuations perfectly match
each other \cite{Mazin2008}.

 \begin{figure}
 \includegraphics[width=\columnwidth]{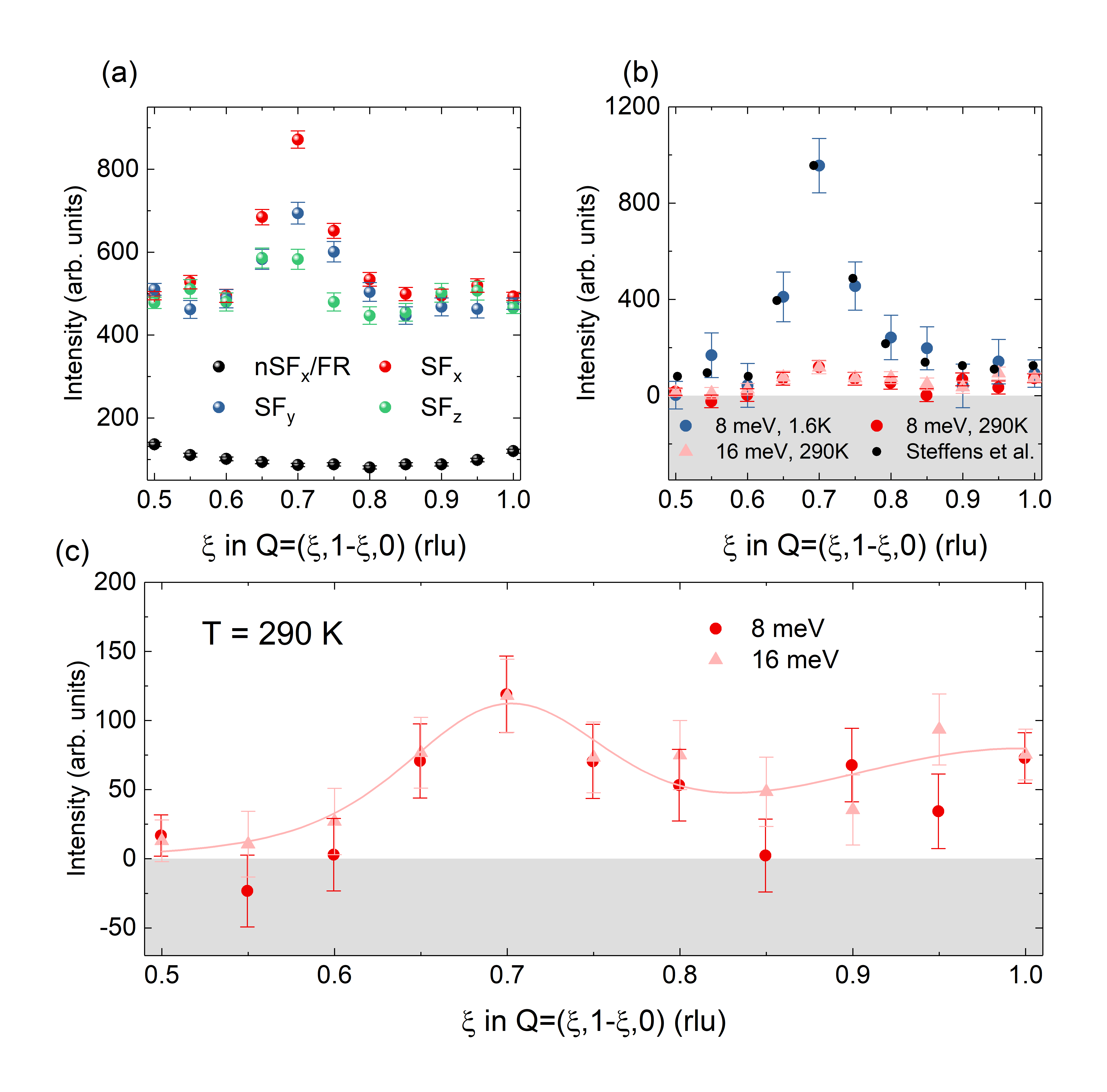}
  \caption{Polarized neutron analysis of the scattering along the diagonal of the 1$^{st}$ Brillouin zone. (a) Example of constant energy scans for all three spin flip channels displays the increased scattering at the incommensurate position (0.7,0.3,0) at 8 meV and 1.6 K. The polarization analysis of all channels yields the purely magnetic scattering signal displayed in (b) for different energies and temperatures. The black circles represent data of the previously reported polarization analysis, taken from \cite{Steffens2019}. This data set was as well measured at 8 meV and 1.6 K. (c) The magnetic signal at 290 K can be described by the susceptibility model used in \cite{Steffens2019} (light red line).
  The intensity in (a) is normalized with 7800000 monitor counts
which corresponds to a measuring time of about 20 minutes per point.}
  \label{ICpolAna}
 \end{figure}


 \begin{figure}
 \includegraphics[width=0.8\columnwidth]{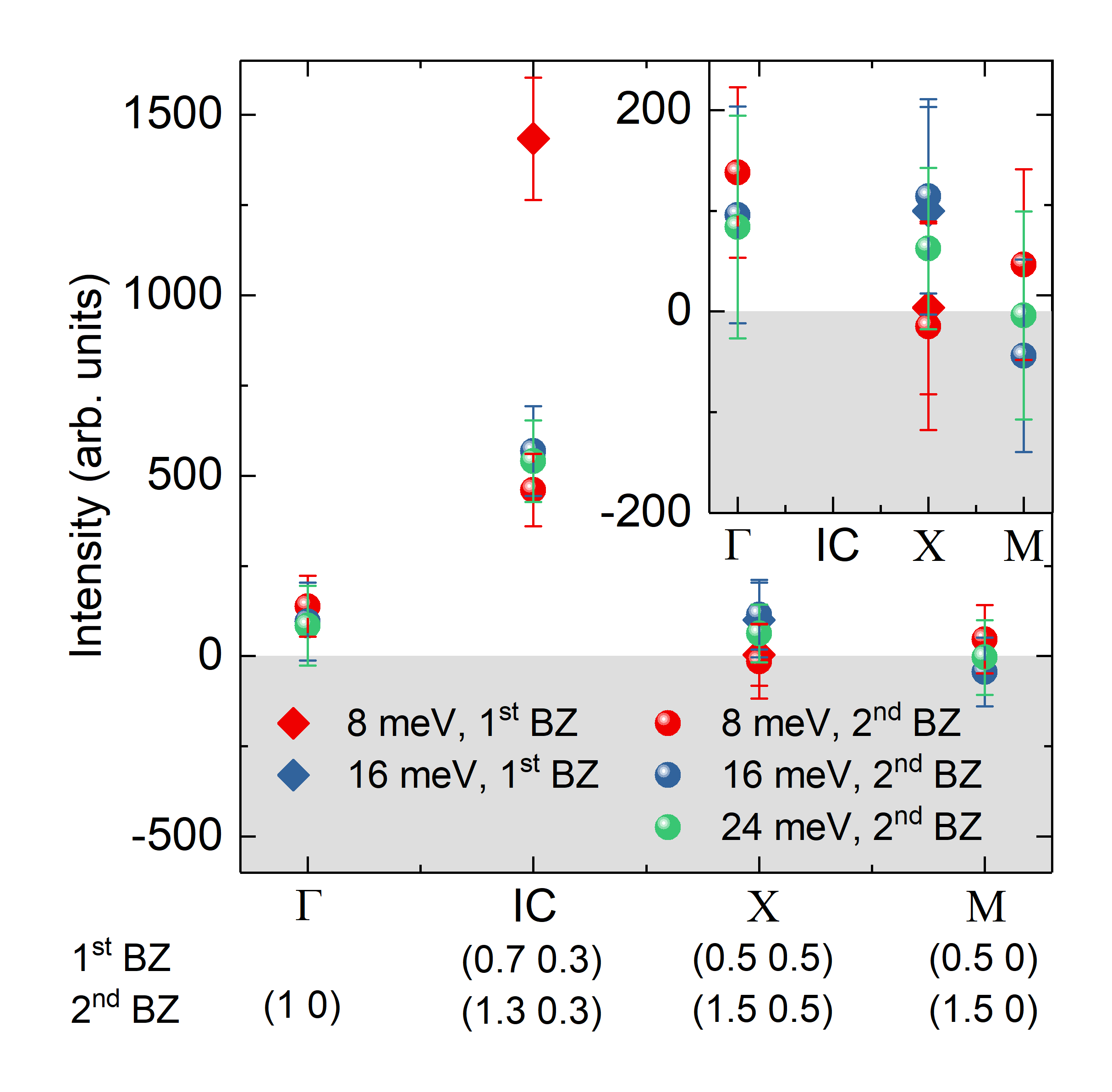}
  \caption{Comparison of magnetic scattering (T=1.6\,K) at prominent points in k-space with $L$=0. The magnetic signal was extracted using the polarization analysis (2I(SF$_x$)-I(SF$_y$)-I(SF$_z$)) and is displayed for the points in the Brillouin zone and different energies: $\Gamma$ point, the incommensurate position, and the different zone boundaries \textrm{X} and \textrm{M}. The inset magnifies the intensity region around zero.
  }
  \label{BZpoints}
 \end{figure}

\subsection{Shape of the quasi-ferromagnetic fluctuations}

The polarization analysis of inelastic neutron scattering provides the separation of the magnetic from any other scattering contribution.
It is therefore possible to identify a tiny magnetic response that is little structured in $\bf q$ space.
This technique was used to detect quasi-ferromagnetic fluctuations and to determine their strength in comparison to the incommensurate fluctuations in \sro \cite{Steffens2019}.
We wished to extend this study focusing on the $\bf q$ dependence of the magnetic quasi-ferromagnetic response.
Recent DMFT calculations \cite{Strand2019} find evidence for local fluctuations superposing the well-established nesting excitations,
which qualitatively agree with the experimental quasi-ferromagnetic signal.
However, while
the neutron experiments indicate a finite suppression of the quasi-ferromagnetic response towards the boundaries of the Brillouin zone, the DMFT
calculation obtains an essentially local feature without such $\bf q$ dependence.

The polarized neutron study was performed on the thermal TAS IN20 and the results are shown in Fig.  \ref{ICpolAna}. An example of the raw data with different spin channels that are needed for the polarization analysis, is given in Fig. \ref{ICpolAna}(a) where a diagonal constant energy scan at 8\,meV, reaching from the zone boundary (0.5,0.5) over the incommensurate position (0.7,0.3) to the zone center (1,0), is shown.
The $x$, $y$, $z$ indices refer to the common coordinate system used in neutron polarization analysis in respect to the scattering vector $\mathbf{Q}$ \cite{Steffens2019}.
The three spin flip channels SF$_x$, SF$_y$, and SF$_z$ clearly exhibit a maximum at the incommensurate position.
While SF$_y$ and SF$_z$ exhibit comparable amplitudes SF$_x$ carries the doubled intensity as it senses both magnetic components perpendicular to the scattering vector.
There is an enhancement of magnetic excitations polarized along the $c$ direction, that can be seen in the stronger SF$_y$ and that was studied in reference \onlinecite{Braden2004}.
Assuming a polarization independent background, 2I(SF$_x$)-I(SF$_y$)-I(SF$_z$) yields the background free magnetic signal, see discussion in reference \onlinecite{Steffens2019}.

Fig. \ref{ICpolAna}(b) displays the magnetic signal, corrected for Bose factor, i.e. the imaginary part of the susceptibility, for different energies and temperatures.
The data well agrees with the results for 8\,meV and 1.6\,K presented in \cite{Steffens2019}.
Additionally, the data at 290\,K indicates a significant drop of the incommensurate nesting signal, which however still is finite and clearly observable.
The temperature dependence of the incommensurate signal was first discussed in reference \onlinecite{Sidis1999}, where the neutron scattering results are compared to the NMR results from reference \onlinecite{Imai1998}. The incommensurate signal was found to strongly decrease with increasing temperature up to room temperature while the ferromagnetic component of the NMR is nearly temperature independent. Also the previous polarized neutron experiment found the quasi-ferromagnetic contribution to be almost identical at 1.6 and 160\,K \cite{Steffens2019}.
As indicated in Fig. \ref{ICpolAna}(b) the quasi-ferromagnetic signal does not change up to 290\,K, so that the peak heights of incommensurate and quasi-ferromagnetic contributions are comparable at ambient temperature.
Taking the much broader $q$ shape of the quasi-ferromagnetic excitations into account, the $q$-integrated spectral weight of the latter clearly dominates.
Around room temperature the quasi-ferromagnetic fluctuations possess thus a larger impact on any integrating processes such as electron scattering.
The quasi-ferromagnetic fluctuations at 290\,K, however, do not exhibit a local character as the signal is significantly reduced at the antiferromagnetic zone boundary (0.5,0.5,0) (Fig. \ref{ICpolAna}(c)). This confirms the conclusion of Steffens et al. \cite{Steffens2019} that the quasi-ferromagnetic fluctuations are sharper in $q$ space than expected from the calculations. Also at the other zone boundary (0.5,0,0) there is no significant magnetic signal detectable (see Fig. \ref{BZpoints}).

\section{Conclusion}

Polarized and unpolarized neutron scattering experiments were performed to study several aspects of the magnetic fluctuations in Sr$_2$RuO$_4$ that are particularly relevant for a possible superconducting pairing scenario.
 The TOF instrument LET yields full mapping of the excitations and reveals the well-studied incommensurate fluctuations at (0.3,0.3) in the two-dimensional reciprocal space.
 There is also ridge scattering at (0.3,$\xi$) reflecting the one-dimensional character of the $d_{xz}$ and $d_{yz}$ bands, as first reported in references \onlinecite{Iida2011,Iida2012}.
 These ridges are stronger between the four peaks surrounding (0.5,0.5), i.e. for $\xi$$>$$0.3$, but the suppression of the signal at smaller $\xi$ is gradual.
 The TOF data confirm the pronounced asymmetry of the nesting peaks.
 Concerning the study of the nesting fluctuations at very low energy in the superconducting phase, TAS experiments yield higher statistics due to the possibility to focus the experiment on the particular position in $\bf Q$,E space.
 Data taken at different out-of-plane components of the scattering vector exclude a sizeable resonance mode emerging at $L$=0.5 in the superconducting phase.
 Only by combining the results of several experiments one can obtain some evidence for the suppression of spectral weight at very low energies.

 With neutron polarization analysis the magnetic excitations were further characterized at 290\,K.
 The incommensurate nesting signal is strongly reduced but still visible, while the quasi-ferromagnetic contribution is almost unchanged.
 At this temperature there is a suppression of this quasiferromagnetic scattering at the Brillouin-zone boundaries, which underlines that this response is not fully local.


We acknowledge stimulating discussions with Ilya Eremin. This work was funded by the Deutsche Forschungsgemeinschaft (DFG,
German Research Foundation) - Project number 277146847 - CRC 1238, project B04, the JSPS KAKENHI
Nos. JP15H05852 and JP17H06136, and the JSPS core-to-core program.

\end{document}